\documentstyle[aps,prl,preprint]{revtex}
\draft
\begin{document}
\title{Generating Converging Eigenenergy
 Bounds for the Discrete States of the  $-ix^3$ Non-Hermitian Potential 
}
\author{C. R. Handy}
\address{Department of Physics \& Center for Theoretical Studies of 
Physical Systems, Clark Atlanta University, 
Atlanta, Georgia 30314}
\date{Received \today}
\maketitle
\begin{abstract}
Recent investigations by Bender and Boettcher (Phys. Rev. Lett 
80, 5243 (1998)) and 
Mezincescu ( J. Phys. A. 33, 4911 (2000)) have argued that the 
discrete spectrum of the non-hermitian potential $V(x) = -ix^3$ should 
be real. We give further evidence for this through a novel formulation
which transforms the general 
 one dimensional Schrodinger equation (with complex potential)
into a fourth order linear differential equation for $|\Psi(x)|^2$. 
This permits the application of the Eigenvalue Moment Method,
developed by Handy, Bessis, and coworkers (Phys. Rev. Lett. 
55, 931 (1985);60, 253 (1988a,b)), 
yielding rapidly converging lower and upper 
bounds to the low lying discrete state energies. We adapt this
formalism to the 
 pure imaginary cubic potential, generating tight
bounds for the first five discrete state energy levels.

\end{abstract}
\vfil\break
\section{Introduction}

In the recent work by Bender and Boettcher (1998) they conjectured that
certain
${\cal P}{\cal T}$ invariant systems should have real discrete spectra.
Various examples were presented, including the $-ix^3$ potential. 
The interest in such systems has increased,  particulary through the
more recent work of Bender et al (1999), Bender et al (2000),
Bender and Wang (2001),  Caliceti (2000), Delabaere and Pham (1999), 
Delabaere and Trinh (2000), Levai and Znojil (2000), Mezincescu (2000, 2001),  
Shin (2000), and Znojil (2000).

We present a radically new way of attacking such 
problems. Although the results presented here combine rigorous mathematical 
theorems and their numerical implementation, it should also be possible to
develop them purely within an algebraic context,
 and confirm that the $-ix^3$
potential can only have real discrete spectra. This  particular approach
is under investigation, and the results will be presented elsewhere. 
However, we have been able to implement the procedure discussed below, 
numerically, for the case of complex energies, $E$, and find no evidence
for such discrete states (for moderate energy values). The details of
this will be communicated in a forthcoming work focusing on the $ix^3 + 
i\alpha  x$ potential studied by Delabaere and Trinh (2000). Our principal
objective in this communication is to emphasize the importance of positivity
as a quantization condition, within the appropriate (moment based) 
representation.

Our starting point is the observation 
 that the one dimensional Schrodinger 
equation (on the real line),
\begin{equation}
-\partial_x^2\Psi(x) + V(x) \Psi(x) = E \Psi(x),
\end{equation}
for complex potentials, $V = V_R + iV_I$, and real energies, $Im E = 0$, can be transformed into
a fourth order, linear differential equation for $S(x) = |\Psi(x)|^2$ :
\begin{eqnarray}
-{1\over {V_I}}S^{(4)} - \Big({1\over{V_I}}\Big)'S^{(3)}
+ 4 \Big( {{V_R-E}\over {V_I}} \Big) S^{(2)}
+ \Big( 4\Big({{ V_R}\over { V_I}}\Big)'+ 2\Big({{{V_R}'}\over
{ V_I}}\Big )
-4E \Big({1\over{ V_I}}\Big)' \Big) S^{(1)} \cr
+ \Big( 4  V_I + 2\Big({{{V_R}'}\over { V_I}}\Big)'\Big) S = 0,
\end{eqnarray}
where $S^{(i)} \equiv \partial_x^iS$.
This equation assumes that the eigenenergy, $E$, is real. We derive it
in the next section.

We could also assume 
that $E$ is complex and incorporate its imaginary part into $V_I$. Since our
objective is to show, numerically, that the conjecture that $E$ is real
is a viable one, we restrict our considerations to this case only, here.  
The method presented in this work
  is so powerful (both theoretically and numerically)
that if the discrete state is not purely real, then it will be detected, at
some sufficiently high calculation order.

The above fourth order differential equation can be generalized to include any
complex contour in the complex plane. However, for the particular problem considered
here, we have only focused on the simplest representation for $S(x)$, as 
given by Eq.(2).

If the potential is real, $V_I = 0$, then Handy et al  (1987a,b; 1988c)
 have shown that $S(x)$ 
satisfies a third order differential equation. This is easy to see from the
above by simply taking $V_I \rightarrow 0$, and recognizing that Eq.(2) 
becomes the total derivative of the third order equation

\begin{equation}
-{1\over 2} S^{(3)}(x) + 2V(x) S^{(1)}(x) + V'(x) S(x) = 2E S^{(1)}(x).
\end{equation}

The importance of converting the discrete state problem into the nonnegative 
$S(x)$ representation is that for rational fraction complex potentials, one
can then exploit the Eigenvalue Moment Method (EMM) of Handy, Bessis, 
and coworkers (1985,1988a,b), enabling the generation of converging lower 
and upper bounds for the low lying discrete states.
 
For rational fraction potentials, Eq.(2) can be transformed 
into a moment equation involving the Hamburger moments
\begin{equation}
\mu_p \equiv \int_{-\infty}^{+\infty} dx \ x^p S(x),
\end{equation}
$ p \geq 0$.
The Moment Equation (ME) takes on the form
\begin{equation}
\mu_p = \sum_{\ell = 0}^{m_s} M_{p,\ell}(E) \mu_\ell,
\end{equation}
$p \geq 0$, where the energy dependent coefficients are easily obtained, and 
satisfy (i.e. ``initialization conditions"): $ M_{\ell_1,\ell_2} = \delta_{\ell_1,\ell_2}$, for $0 \leq \ell_{1,2} \leq m_s$. The {\it missing moments}, $\{\mu_\ell| 0 \leq \ell \leq m_s\}$, 
are to be considered as independent variables. The missing moment order, $m_s$,
is problem dependent.

The homogeneous nature of the Schrodinger equation requires the imposition of
an appropriate normalization condition. Although this requires some care, 
usually, a convenient choice is to take 
\begin{equation}
\sum_{\ell = 0}^{m_s} \mu_p = 1.
\end{equation}
Solving for $\mu_0$, and substituting into the ME relation, gives
\begin{equation}
\mu_p = \sum_{\ell = 0}^{m_s} {\hat M}_{p,\ell}(E) {\hat \mu}_\ell,
\end{equation}
where 
\begin{equation}
{\hat \mu}_\ell = \cases { 1, \ \ell = 0 \cr
			  \mu_\ell, \ 1 \leq \ell \leq m_s \cr}
\end{equation}
and
\begin{equation}
{\hat M}_{p,\ell}(E) = \cases { M_{p,0}(E), \ \ell = 0 \cr
			       M_{p,\ell}(E) - M_{p,0}(E), \ 1 \leq \ell \leq m_s \cr} .
\end{equation}

From the Hankel-Hadamard (HH) positivity theorems (Shohat and Tamarkin (1963)), the
Hamburger moments must satisfy the
conditions $\int_{-\infty}^{+\infty}dx \ \Big(\sum_{j = 0}^J C_j x^j)^2 S(x) > 0$, for all $C$'s and $J \geq 0$. These become the quadratic form expressions

\begin{equation}
\sum_{j_{1,2} = 0}^J C_{j_1} \mu_{j_1+j_2} C_{j_2} > 0.
\end{equation}
In terms of the (unconstrained) normalized $\mu$'s this becomes

\begin{equation}
\sum_{\ell = 0}^{m_s} {\hat \mu}_{\ell}
\Big ( \sum_{j_{1,2} = 0}^J C_{j_1} {\hat M}_{j_1+j_2,\ell}(E) C_{j_2} \Big ) > 0,
\end{equation}
which defines the linear programming   equations (Chvatal (1983)):

\begin{equation}
\sum_{\ell = 1}^{m_s} {\cal A}_\ell(C,E) \mu_\ell < {\cal B}(C,E),
\end{equation}
for all possible $C$'s (except those identically zero), where
\begin{equation}
{\cal A}_\ell(C,E) = -\Big ( \sum_{j_{1,2} = 0}^J C_{j_1} {\hat M}_{j_1+j_2,\ell}(E) C_{j_2} \Big ),
\end{equation}
and 
\begin{equation}
{\cal B}(C,E) = 
\Big ( \sum_{j_{1,2} = 0}^J C_{j_1} {\hat M}_{j_1+j_2,0}(E) C_{
j_2} \Big ).
\end{equation}

If at a given order, $J$, and arbitrary energy value, $E$, there exists a solution set to all of the above inequalities, ${\cal U}_E^{(J)}$,
 then it must be convex. Through a linear programming based {\it cutting} procedure (Handy et al (1988a,b)),
 one can find optimal $C$'s which (in a finite number of steps) establish
the existence or nonexistence of ${\cal U}_E^{(J)}$. The energy values
for which  missing moment solution sets exist, define energy intervals,

\begin{equation}
E \in \bigcup_{n = 0}^{N(J)} [E_{L;n}^{(J)},E_{U;n}^{(J)}], \ {\rm {if}}\ {\cal U}_E^{(J)}  \neq \oslash,
\end{equation}
which become smaller as $J$ increases, converging to the corresponding discrete 
state energy (which must always lie within the respective interval):

\begin{equation}
 E_{L;n}^{(J)} \leq E_{L;n}^{(J+1)}  \leq \ldots \leq E_{physical;n}
\leq \ldots \leq  E_{U;n}^{(J+1)} \leq E_{U;n}^{(J)} .
\end{equation}
Through the EMM approach, we can easily generate the converging lower and upper 
bounds to the desired discrete state energy.

We note that although the traditional Moment Problem theorems are concerned with uniqueness questions (i.e. is there a unique function with the moments $\mu_p$ 
satisyfing the HH positivity conditions ?), within the context of physical systems such issues are usually inconsequential. This is because the very nature of
the ME relation will guarantee uniqueness. That is, our moments are associated
with an underyling differential equation with unique physical solutions.
\vfil\break
\section{ Deriving the Positivity Equation for $S(x)$ }

We derive Eq.(2) as follows. First, multiply the
Schrodinger equation ($E$ real) by $\Psi^*$ :
\begin{equation}
-\Psi^*(x)\Psi''(x) +  V(x) S(x) = E S(x).
\end{equation}
The complex conjugate becomes
\begin{equation}
-\Psi(x){\Psi^*}''(x) +  V^*(x) S(x) = E S(x).
\end{equation}
Adding both expressions, and using $\Psi^* \Psi'' = (\Psi^*\Psi')'-|\Psi'|^2$, yields 
\begin{equation}
-[S'' - 2|\Psi'|^2] + 2 { V}_R S = 2 E S.
\end{equation}
This in turn becomes (upon differentiating)
\begin{equation}
-S''' + 2\Big(|\Psi'|^2\Big)' + 2 \Big({ V}_R S\Big)' = 2 E S'.
\end{equation}

If we subtract Eq.(18) from Eq.(17), then
\begin{equation}
\partial_x\Big({\Psi^*} \Psi' - \Psi {\Psi^*} ' \Big ) = 2i { V}_I S.
\end{equation}

Returning to the Schrodinger equation, we multiply both sides by ${\Psi^*}'$:
\begin{equation}
-{\Psi^*}'\Psi'' + { V} \Psi {\Psi^*}' = E \Psi {\Psi^*}'.
\end{equation}
The complex conjugate is
\begin{equation}
-\Psi'{\Psi^*}'' + {{ V}^*} {\Psi^*} \Psi' = E {\Psi^*} \Psi'.
\end{equation}
Substituting ${ V} = { V}_R + i{ V}_I$, we add both expressions (and divide by $i{ V}_I$):
\begin{equation}
-{{\Big({|\Psi'|^2}\Big )'}\over {i{ V}_I}} + {{{ V}_R S'}\over {i{ V}_I}}
 +  [\Psi{\Psi^*}'-{\Psi^*}\Psi'] = E {{S'}\over {i{ V}_I}}.
\end{equation}
Differentiating with respect to $x$, and substituting Eq.(21) yields
\begin{equation}
-\Big({{\big({|\Psi'|^2}\big )'}\over {i{ V}_I}}\Big)' + \Big({{{ V}_R S'}\over {i{ V}_I}}\Big)'
 -2i{ V}_I S = E \Big({{S'}\over {i{ V}_I}}\Big)'.
\end{equation}

Upon dividing Eq.(19) by  $i{ V}_I$, and differentiating, we obtain
\begin{equation}
-\Big({{S'''}\over {i{ V}_I}}\Big)' + 2\Big({{\big(|\Psi'|^2\big)'}\over  {i{ V}_I}}
\Big )'
 + 2 \Big({{\big({ V}_R S\big)'}\over {i{ V}_I}}\Big)' = 2E \Big({{S'}\over {i{ V}_I}}\Big)'.
\end{equation}
Finally, we substitute Eq.(25) for the second term in Eq.(26), obtaining a fourth order
linear differential equation for $S$:
\begin{equation}
-\Big({{S'''}\over {i{ V}_I}}\Big)'
+ 2 \times \Big ( \big({{{ V}_R S'}\over {i{ V}_I}}\big)' -2i{ V}_I S -E \big({{S'}\over {i{ V}_I}}\big)'\Big ) + 2 \Big({{\big({ V}_R S\big)'}\over {i{ V}_I}}\Big)' = 2E \Big({{S'}\over {i{ V}_I}}\Big
)',
\end{equation}
or
\begin{equation}
-\Big({{S'''}\over {{ V}_I}}\Big)'
+ 4 \times \Big ( \Big({{{ V}_R S'}\over {{ V}_I}}\Big)' + { V}_I S 
\Big ) + 2 \Big({{{ V}_R' S}\over {{ V}_I}}\Big)' = 4E \Big({{S
'}\over {{ V}_I}}\Big
)',
\end{equation}
which becomes Eq.(2).

The positivity differential representation in Eq.(2) is a fourth order linear differential equation, with four independent solutions, for any $E$.  Within the EMM formalism,
it is important to prove that the physical solution is the only one which is
both nonnegative ($S(x) \geq 0$) and bounded, with finite moments (i.e. $S(x)$ is 
in $L^2$). We can prove this for Eq.(2).

For any  real energy variable value, $E \in \Re$,
 let $\Psi_1(x)$ and $\Psi_2(x)$
denote the two independent solutions to the Schrodinger equation.
The expression $S(x) = |\alpha \Psi_1(x) + \beta \Psi_2(x)|^2 = 
|\alpha|^2 \times |\Psi_1(x)|^2 + |\beta|^2\times |\Psi_2(x)|^2 + 
\alpha \beta^*\Psi_1(x)\Psi_2^*(x) + \alpha^*\beta \Psi_1^*(x)\Psi_2(x)$,  then
becomes a solution to Eq.(2).
So too are $|\Psi_1(x)|^2$ and $|\Psi_2(x)|^2$.
Accordingly, since $\alpha$ and $\beta$ are arbitrary, and $\Psi_1(x)$ and
$\Psi_2(x)$ are complex, the configurations $\Psi_1(x)\Psi_2^*(x)$ and
$\Psi_1^*(x)\Psi_2(x)$ are independent (complex) solutions to Eq.(2) as
well. 

From low order JWKB asymptotic analysis (Bender and Orszag (1978)), 
in either asymptotic direction ($x \rightarrow \pm \infty$), one of the
semiclassical modes will be exponentially increasing, while the other
is exponentially decreasing. Therefore it becomes clear that the only
 possible nonnegative and bounded $S(x)$ configuration is that 
corresponding to the physical solutions.

\vfil\break
\section{ The $-ix^3$ Potential }

The positivity differential equation for the $V(x) = -ix^3$ potential is
(i.e. $V_R = 0, V_I = -x^3$)
 
\begin{eqnarray}
x^{-3}S^{(4)}(x) - 3 x^{-4} S^{(3)}(x)
+ 4 E x^{-3} S^{(2)}(x)
-12Ex^{-4} S^{(1)}(x) 
-4 x^3 S(x) = 0.
\end{eqnarray}
Multiplying both sides by $x^{p+4}$, and integrating over $\Re$,
 produces the ME relation
\begin{equation}
4\mu_{p+7} = (p+4)p(p-1)(p-2)\mu_{p-3}
 +   4Ep(p + 4)  \mu_{p-1},
\end{equation}
for $p \geq 0$.

The moment equation separates into two relations, one for the odd moments, the other for the even moments.  Assuming that the discrete states are 
nondegenerate and have real eigenenergies, we have:

\begin{equation}
\Psi^*(-x) = \Psi(x),
\end{equation}
and 
\begin{equation}
S(-x) = \Psi^*(-x) \Psi(-x) = \Psi(x) \Psi^*(x) = S(x).
\end{equation}
Thus, the physical $S(x)$'s are symmetric, and the odd order moments are zero. 

The even order Hamburger moments
\begin{equation}
\mu_{2\rho} \equiv u_\rho,
\end{equation}
correspond to the Stieltjes moments,
\begin{equation}
u_\rho \equiv \int_0^\infty dy \ y^\rho \Upsilon(y),
\end{equation}
of the function
\begin{equation}
\Upsilon (y) \equiv {{S(\sqrt {y})}\over {\sqrt {y}}}.
\end{equation}
The corresponding Stieltjes moment equation for the $-ix^3$ potential becomes
(i.e. substitute $p = 2\rho+1$ in Eq.(30))
\begin{equation}
4u_{\rho+4} = (2\rho+5)(2\rho+1)(2\rho)(2\rho-1)u_{\rho-1}
 +   4E(2\rho+1)(2\rho+5)  u_{\rho},
\end{equation}
for $\rho \geq 0$. This is an $m_s = 3$ order problem. One can convert this 
into the form in Eq.(5) (i.e. $u_\rho = \sum_{\rho = 0}^{m_s}M_{\rho,\ell}(E) 
u_\ell$), where the $M$ coefficients satisfy Eq.(36), with respect to the
first index ($\rho$), as well as the initial conditions previously identified.

 One convenient feature about the Stieltjes representation is that the normalization condition

\begin{equation}
\sum_{\ell  = 0}^3 u_\ell = 1,
\end{equation}
involves nonnegative moments.

From the Stieltjes moment problem (Shohat and Tamarkin (1963)) we know that
the counterpart to Eq.(10) is

\begin{equation}
\sum_{j_{1,2} = 0}^ J C_{j_1} u_{\sigma + j_1 + j_2} C_{j_2} > 0,
\end{equation}
for $\sigma = 0, 1$. Accordingly, the necessary linear programming equations
to consider are

\begin{equation}
\sum_{\ell = 1}^{m_s} {\cal A}_{\ell}(C,E;\sigma) < {\cal B}(C,E;\sigma),
\end{equation}
where 
\begin{equation}
{\cal A}_\ell(C,E;\sigma) = -\Big ( \sum_{j_{1,2} = 0}^J C_{j_1} {\hat M}_{\sigma + j_1+j_2,\ell}(
E) C_{j_2} \Big ),
\end{equation}
and
\begin{equation}
{\cal B}(C,E;\sigma) =
\Big ( \sum_{j_{1,2} = 0}^J C_{j_1} {\hat M}_{\sigma+j_1+j_2,0}(E) C_{
j_2} \Big ).
\end{equation}

The numerical implementation of the EMM procedure yields the excellent results
quoted in Tables I - V. 
Our results are in agreement with those of Bender and Boettcher (1998), 
as well as those of Handy, Khan, and Wang (2000). We indicate the maximum moment order generated, ${P}_{max}$,
through the ME relation.

Since our results are based on  equations that
explicitly assume $E$ is real, and the EMM procedure is very stable and 
highly accurate (as evidenced through the tightness of its bounds), any
imaginary part to the discrete state energy would reveal itself through 
some anomalous behavior in the generated bounds. That is, at some
order $P_{max}$, no feasible energy interval would survive (i.e. 
${\cal U}_E^{(J)} = \oslash$, for all $E$). This is never observed,
to the order indicated. As such, our analysis strongly supports the
reality of the (low lying) discrete state spectrum for the $-ix^3$
potential.

\begin{table}
\caption {Bounds for the Ground State Energy of the $-ix^3$ Potential}
\begin{center}
\begin{tabular}{cccccl}
 \multicolumn{1}{c}{$P_{max}$}
& \multicolumn{1}{c}{$E_{L;0}$} & \multicolumn{1}{c}{$E_{U;0}
$}\\ \hline
10 & .825  &        1.405 \\
20 & 1.15619 &        1.15645 \\
30 & 1.1562669 &        1.1562672 \\
40 & 1.1562670718 &        1.1562670721 \\
50 & 1.156267071988016 &   1.156267071988161 \\
60 & 1.15626707198811324 &   1.15626707198811335
\end{tabular}
\end{center}
\end{table}

\begin{table}
\caption {Bounds for the First Excited State Energy of the $-ix^3$ Potential}
\begin{center}
\begin{tabular}{cccccl}
 \multicolumn{1}{c}{$P_{max}$}
& \multicolumn{1}{c}{$E_{L;1}$} & \multicolumn{1}{c}{$E_{U;1}
$}\\ \hline
20 &         4.1056  &        4.1168 \\
30 &         4.109225  &        4.109236 \\
40 &         4.1092287509  &        4.1092287578 \\
50 &         4.109228752806  &        4.109228752812 
\end{tabular}
\end{center}
\end{table}

\begin{table}
\caption {Bounds for the Second Excited State Energy of the $-ix^3$ Potential}
\begin{center}
\begin{tabular}{cccccl}
 \multicolumn{1}{c}{$P_{max}$}
& \multicolumn{1}{c}{$E_{L;2}$} & \multicolumn{1}{c}{$E_{U;2}
$}\\ \hline
20 &         7.420     &    7.594 \\
30 &         7.56213     &    7.56242 \\
40 &         7.562273794     &    7.562273999 \\
50 &         7.5622738549    &     7.5622738551
\end{tabular}
\end{center}
\end{table}

\begin{table}
\caption {Bounds for the Third Excited State Energy of the $-ix^3$ Potential}
\begin{center}
\begin{tabular}{cccccl}
 \multicolumn{1}{c}{$P_{max}$}
& \multicolumn{1}{c}{$E_{L;3}$} & \multicolumn{1}{c}{$E_{U;3}
$}\\ \hline
30 &         11.3115  &        11.3159 \\
40 &         11.314418  &        11.314425 \\
50 &         11.314421818  &        11.314421824
\end{tabular}
\end{center}
\end{table}

\begin{table}
\caption {Bounds for the Fourth Excited State Energy of the $-ix^3$ Potential}
\begin{center}
\begin{tabular}{cccccl}
 \multicolumn{1}{c}{$P_{max}$}
& \multicolumn{1}{c}{$E_{L;4}$} & \multicolumn{1}{c}{$E_{U;4}
$}\\ \hline
30 & 15.20  &        15.80 \\
40 & 15.29145  &         15.29160 \\
50 & 15.29155366 &        15.29155380 \\
60 & 15.29155375037  &        15.29155375041 
\end{tabular}
\end{center}
\end{table}

\vfil\break
\section {Acknowledgments}

This work was supported in part by a grant from the National Science Foundation
(HRD 9632844) through the Center for Theoretical Studies of Physical Systems (CTSPS). The author is appreciative of stimulating discussions with Dr. Alfred Z. Msezane, Dr. G.  Andrei Mezincescu, and Dr. Daniel Bessis, as well as comments
by Dr. Carl Bender,
which impacted this work.
\vfil\break
\section {References}
\noindent Bender C M and Boettcher S 1998 Phys. Rev. Lett. {\bf 80} 5243

\noindent Bender C M, Boettcher S, and Meisinger P N 1999, J. Math. Phys. {\bf 40} 2201

\noindent Bender C M, Boettcher S, Jones H F and Savage V M 1999 J. Phys. A:
Math. Gen. {\bf 32} 1

\noindent Bender C M, Boettcher S and Savage V M 2000 J. Math. Phys. {\bf 41} 6381

\noindent Bender C M and Wang Q 2001 J. Phys. A: Math. Gen.  

\noindent Bender C M and Orszag S A,   {\it Advanced Mathematical
Methods for
Scientists and Engineers} (New York: McGraw Hill 1978).

\noindent Caliceti E 2000 J. Phys. A: Math. Gen. {\bf 33} 3753

\noindent Chvatal V 1983 {\it Linear Programming} (Freeman, New York).

\noindent Delabaere E and Pham F 1998 Phys. Lett. {\bf A250} 25

\noindent Delabaere E and Trinh D T 2000 J. Phys. A: Math. Gen. {\bf 33} 8771

\noindent Handy C R 1987a Phys. Rev. A {\bf 36}, 4411 

\noindent Handy C R 1987b Phys. Lett. A {\bf 124}, 308 

\noindent Handy C R and  Bessis D 1985 Phys. Rev. Lett. {\bf 55}, 931 

\noindent Handy C R, Bessis D, and Morley T D 1988a Phys. Rev. A {\bf 37}, 4557 

\noindent Handy C R, Bessis D, Sigismondi G, and Morley T D 1988b Phys. Rev. Lett
{\bf 60}, 253 

\noindent Handy C R, Khan D, and Xian Qiao Wang 2001 CAU preprint

\noindent Handy C R, Luo L, Mantica G, and Msezane A 1988c Phys. Rev. A {\bf 38}, 490

\noindent Levai G and Znojil M 2000 J. Phys. A: Math. Gen. {\bf 33} 7165

\noindent Mezincescu G A 2001 J. Phys. A: Math. Gen.

\noindent Mezincescu G A 2000 J. Phys. A: Math. Gen. {\bf 33} 4911

\noindent Shohat J A and  Tamarkin J D, {\it The Problem of Moments}
(American Mathematical Society, Providence, RI, 1963).

\noindent Shin K C 2000 Preprint math-ph/0007006 (J. Math. Phys, under press)

\noindent Znojil M 2000 J. Phys. A: Math. Gen. {\bf 33} 6825

\end{document}